\documentclass[%
 reprint,
superscriptaddress,
 amsmath,amssymb,
 aps,
onecolumn,
]{revtex4-2}

\usepackage{graphicx}
\usepackage{dcolumn}
\usepackage{bm}
\usepackage{physics}
\usepackage{xcolor}
\usepackage{comment}

\begin{document}

\newcommand{\az}[1]{{\color{blue}#1}} 
\newcommand{\dm}[1]{{\color{magenta}#1}} 
\newcommand{\rev}[1]{#1} 
\newcommand{\lmax}{L_\mathrm{max}}
\newcommand{\pmax}{\phi_\mathrm{max}}
\newcommand{\atanTwo}{\mathop{\mathrm{atan2}}\nolimits}

\preprint{APS/123-QED}

\title{Optimal Undulatory Swimming with\\Constrained Deformation and Actuation Intervals}

\author{Fumiya Tokoro}
\affiliation{%
  Graduate School of Engineering Science, The University of Osaka
}%

\author{Hideki Takayama}
\affiliation{%
  Graduate School of Engineering Science, The University of Osaka
}%

\author{Shinji Deguchi}
\affiliation{%
  Graduate School of Engineering Science, The University of Osaka
}%

\author{Andreas Z\"{o}ttl}
\email{andreas.zoettl@univie.ac.at}
\affiliation{
  Faculty of Physics, University of Vienna
}%
\author{Daiki Matsunaga}
\email{daiki.matsunaga.es@osaka-u.ac.jp}
\affiliation{%
  Graduate School of Engineering Science, The University of Osaka
}%

\date{\today}

\begin{abstract}
In nature, many unicellular organisms are able to swim with the help of beating filaments, where local energy input leads to cooperative undulatory beating motion.
Here, we investigate by employing reinforcement learning how undulatory microswimmers
\rev{modeled as a discretized bead-bend-spring filament actuated by torques which are constrained locally.}
We show that the competition between actively applied torques and intrinsic bending stiffness leads to various optimal beating patterns characterized by distinct frequencies, amplitudes, and wavelengths.
Interestingly, the optimum solutions depend on the action interval, i.e.\ the time scale how fast the microswimmer can \rev{change the applied torques} based on its internal state.
\rev{We show that optimized stiffness- and action-interval-dependent beating is realized by bang-bang solutions of the applied torques with distinct optimum time-periodicity and phase shift between consecutive joints, which we analyze in detail by a systematic study of possible bang-bang wave solution patterns of applied torques.}
Our work not only sheds light on how efficient beating patterns of biological microswimmers can emerge based on internal and local constraints, but also offers actuation policies for potential artificial elastic microswimmers.
\end{abstract}

\maketitle


\section{Introduction}

Many microorganisms, such as unicellular bacteria or protists, are able to swim through viscous fluids, e.g.\  water.
At the micron scale, the physics is governed by low-Reynolds-number hydrodynamics, where inertia cannot be used for self-propulsion.
In contrast, in order to move forward, microswimmers have to perform non-reciprocal periodic deformations of their shape \cite{Lauga2009a,Elgeti2015,Zottl2016}.
While some organisms and cells, such as amoeba may deform their entire body to move \cite{Barry2010a}, or bacteria swim with the help of rotating quasi-rigid helical flagella attached to their cell body \cite{Lauga2016}, many small eukaryotic organisms use thin waving filaments, known as flagella or cilia \cite{Lighthill1976,Brennen1977}.
These filaments actively deform in a periodic manner with the help of molecular motors running along them, inducing periodic non-reciprocal shape deformations in the form of an (almost) planar travelling wave.
Canonical examples are the motion of flagella attached to sperm cells \cite{Gaffney2011}, or the beating pattern of small swimming worms such as \textit{C.~elegans} \cite{Berman2013}.

From the theoretical point of view, travelling wave solutions had first been addressed by Taylor by considering the swimming of an infinitely extended waving sheet \cite{Taylor1951}.
In the simple case of a pure sinusoidal deformation, this is quantified by a fixed wavelength $\lambda$, wave amplitude $b$, and wave speed $U=\omega / k$, where $\omega = 2\pi / T$ with $\omega$ and $T$ being the angular frequency and period of oscillation, respectively.
For sufficiently small amplitudes $\bar{b} = b/ \lambda \ll 1$, the speed of the Taylor sheet can then be calculated in lowest order to $V=\pi \omega b^2 / \lambda$.
When considering large amplitudes and thus higher orders in $\bar{b}$ \cite{Taylor1951,Sauzade2011}, for a given wave speed, the maximum velocity is then reached for $\bar{b} \approx 0.3$.
Similarly, when thin one-dimensional filaments waving in a plane are considered, the swimming speed still scales as $\sim \omega b^2 / \lambda$, yet the prefactor is smaller compared to the waving sheet and depends on the filament thickness. 
For example, within the framework of resistive force theory, the prefactor depends on the ratio of perpendicular to parallel friction coefficients of the filament locally approximated by a slender rod \cite{Lauga2009a}.

In the past, the question had been addressed if and how the motion patterns of filaments are optimized, for example with respect to maximizing the (hydrodynamic/swimming) efficiency, or by considering energetic contributions.
Indeed it had been demonstrated that propulsion by traveling waves is an optimal strategy for planar beating of filaments \cite{Pironneau1974,Alouges2019a,Lauga2020o}.
While assuming a sinusoidal wave shape is the simplest case to consider, real shapes of sperm flagella or swimming worms deviate from simple sinusoidal shapes.
In fact, it has been shown that the optimal traveling wave forms are not sinusoidal but sawtooth-like \cite{Pironneau1974,Lighthill1976,Tam2011,Spagnolie2010a,Koehler2012,Montenegro-Johnson2014}.
Furthermore, adding a cell body attached to a waving flagellum results in optimal waves with increasing amplitude, similar to what is observed for biological sperm cells \cite{Dresdner1980,Tam2011}.

From a numerical perspective, simple approaches to approximate a waving filament are either by implementing a sequence of $N$ short slender rods with anisotropic friction connected by $N-1$ links, or by a chain of $N$ hydrodynamically interacting spheres \cite{Berman2013}.
A simple examples is a 3-link swimmer (\textit{Purcell swimmer}) \cite{Purcell1977}
where its most efficient periodically varying conformations can be found by optimization \cite{Becker2003,Tam2007}.
The aforementioned studies focused on pure hydrodynamic considerations, lining the stroke kinematics to an optimization problem.
However, bending of the filament typically costs some internal energy.
\rev{Few previous works combined hydrodynamic and elastic effects in actively waving filaments focused on optimizing the swimming efficiency, where frequency and wavelength are fixed \cite{Spagnolie2010a,Lauga2013,Ishimoto2016},
and typically the relevant dimensionless number compares the wavelength to an intrinsic elasto-viscous lengthscale \cite{Wiggins1998,Lagomarsino2003,Gauger2006}.}
\rev{Noteworthy, } considering the effects of bending energy in the optimization problem changes the optimum shapes of waving filaments from a sawtooth-like shape to smoother profiles \cite{Spagnolie2010a,Lauga2013}.

Recently, reinforcement learning (RL) has been applied to
\rev{optimize trajectories of microswimmers in different environments, for example to identify optimal paths when navigating in fluid flow (see e.g.\ \cite{Colabrese2017,Alageshan2020}). So far in most of these attempts the
microswimmers are assumed to be active agents of fixed (e.g.\ spherical) shape and their orientations and velocities are simply assumed to change directly, i.e.\ neglecting the underlying non-reciprocal shape deformations and reorientation mechanisms of real microorganisms.
So far only a few RL studies accounted for explicit and dynamic body shape deformations coupled to the fluid,}
based on information of its internal state (e.g.\ shape)  and/or its environment \rev{(e.g.\ chemical field or fluid flow)},
as initially proposed for high-Reynolds number locomotion \cite{Verma2018}.
These shape changes can either be induced directly as a state in a low-dimensional shape space \cite{Tsang2020}, or by proposing instantaneous forces on the body parts, allowing for continuous shape changes \cite{Hartl2021x}.
These approaches had first been applied to the simple three-bead Najafi Golestanian swimmer moving in 1D \cite{Tsang2020,Hartl2021x}, and extended to $N$-bead swimmers with a larger number of beads $N>3$  \cite{Tsang2020,Jebellat2024,Hartl2025}, up to $N \sim 100$  \cite{Hartl2025}.
As a similar bead-based approach has been extended to 2D, for example, triangular microswimmers consisting of three hydrodynamically interacting beads, which are e.g.\ actuated by pairwise arm forces between the beads \cite{Liu2021c,Bulusu2025} or a one-hinge swimmer with variable arm length \cite{Zou2022}.
Besides these popular bead-based models, resistive force theory \cite{Qin2023c} and the boundary element method \cite{Xiong2024} have been employed.
Finally, RL of shape-deforming microswimmers has been applied to study chemotaxis \cite{Hartl2021x,Xiong2024}, context detection \cite{Zou2024}, cooperative motion \cite{Liu2023}, and undulatory motion in external flow \cite{ElKhiyati2023}.

\rev{So far} previous work on \rev{optimizing} shapes for waving filaments 
\rev{has maximized the corresponding objective function such as the stroke-averaged swimming efficiency}
without setting local constraints.
In biological filaments, however, local constraints along the filament may exist, such as a maximum local energy input, which is realized by dynein motors used to bend flagella and cilia.
Therefore, we introduce here a minimal model that considers activity-induced bending effectively as a locally constrained active bending torque.
This we realize by
\rev{setting bounds on the actuation torques}
applied to \rev{the joints of a} discretized filament, which consists of hydrodynamically interacting beads.
We then apply RL to optimize the
torque pattern along the filament
\rev{which only depend on its instantaneous internal state in order}
to maximize \rev{its swimming speed.}
Interestingly, depending on two relevant dimensionless parameters of the system,  we observe a large variety of emerging dynamic shapes and oscillation frequencies, while the overall swimming speed does not vary tremendously for a fixed maximum local bending torque.
We thus demonstrate that, in contrast to conventionally used predefined frequencies and amplitudes in swimming optimization procedures, they emerge as properties resulting from the competition of three different time scales associated with active bending, passive bending, and internal state perception, respectively.
\rev{We further show that the emerging optimal shapes for different bending stiffness are realized by bang-bang solutions for the torques, i.e.\ they periodically switch between the allowed maximum and minimum values. 
By performing additional systematic simulations of bang-bang torque wave solutions we show that for a given bending stiffness an optimum torque-switching frequency and phase difference between the torques along the filament exists, which can be realized for sufficiently small action interval, i.e.\ the time scale how fast the microswimmer can change its applied torques based on its internal state. In the case of large action intervals the found strategies for the microswimmer are still good but interestingly they come with a different optimum torque frequency and phase difference.}
\rev{All together,} our results are relevant in understanding optimized stroke patterns of planar-wave microswimmers, such as sperm cells or \textit{C.\ elegans} worms moving in viscous fluids, and offer a new model for viscosity-dependent shapes of undulatory microswimmers \cite{Smith2009a,Pierce2025}.

\section{Governing equations and methods}
\subsection{Microswimmer model}
We approximate here the shape of a planar beating microswimmer by simply considering $N$ connected beads of radius $a$ located at positions $\bm{r}_i$ with velocities $\bm{v}_i$, $i=1,\dots,N$.
Bead motions are induced by forces $\bm{f}_i$ acting on them, with the constraint of the total vanishing forces and torques on the swimmer.
These forces are induced by passive and active contributions.
First, the beads are connected by applying a harmonic spring potential, 
\begin{equation}
  U_s =  \frac{1}{2} k_s \sum_{i=1}^{N-1} \left( |\Delta \bm{r}_{i}| - \ell_0 \right)^2 \quad ,   
\end{equation}
where $\Delta \bm{r}_{i} = \bm{r}_{i+1} -  \bm{r}_{i}$, $\ell_0$ is the equilibrium length,
and $k_s$ is the relatively stiff spring constant keeping the inter-beads distance close to $\ell_0$,
and hence the total filament length $\ell$ approximately constant, $\ell \approx (N-1)\ell_0$.
Second, we employ a bending potential that tends to align three consecutive beads,
\begin{equation}
U_b = \frac{1}{2} k_b \sum_{i=1}^{N-2} \left( \theta_{i} - \pi \right)^2 
\end{equation}
where $\theta_{i} = \atanTwo \left(- \Delta \bm{r}_i \times \Delta \bm{r}_{i+1} , - \Delta \bm{r}_i \cdot \Delta \bm{r}_{i+1} \right)$
is the bending angle between a joint formed by three beads with  $0 < \theta_i \le 2\pi$.
Together with the spring potential, the bending potential will lead to a straight equilibrium configuration of the filament.
Finally, in order to allow the filament to actively deform and to swim, we introduce an energy term associated with active bending torques, again acting between three consecutive beads,
\begin{equation}
  U_a =  \sum_{i=1}^{N-2} L_i \theta_{i} 
\end{equation}
where the time-dependent active bending torques $L_i(t)$ are chosen to be 
constrained to 
$-L_\mathrm{max} \leq L_i(t) \leq +L_\mathrm{max}$,
where $L_\mathrm{max}$ is fixed.
Note, in contrast to the passive contributions, $U_a$ is not only a function of the bead positions $\mathbf{r}_i$, but depends \rev{explicitly} on the time-dependent torques $L_i$.
A positive torque ($L_i > 0$) applies forces to reduce the joint angle $\theta_{i}$, while a negative torque ($L_i < 0$) increases $\theta_{i}$.
Hence, by applying time-dependent active torques $L_i(t)$ the filament is able to actively deform, and specific choices of  $L_i(t)$ will lead to specific swimming patterns, which we will optimize by reinforcement learning, as outlined below in section \ref{sec:rl}.
The active torque is where the energy is put into the system while maintaining an overall force- and torque-free condition.
In contrast to previous optimization procedures used for microswimmers, which \rev{optimize} for efficiency or minimal energy consumption, our approach thus sets a local constraint on the \rev{torque} input  given by the value of $\lmax$,
which in addition limits the total power consumption of the swimmer (see below).
The total potential of the system $U$ is then given by
\begin{equation}
  U = U_s + U_p + U_a \quad,
\end{equation}
and the force $\bm{f}_i$ applied to each bead reads
\begin{equation}
  \bm{f}_i = - \pdv{U}{\bm{r}_i} \quad .
\end{equation}

\begin{figure}
    \centering
    \includegraphics[width=0.75\columnwidth]{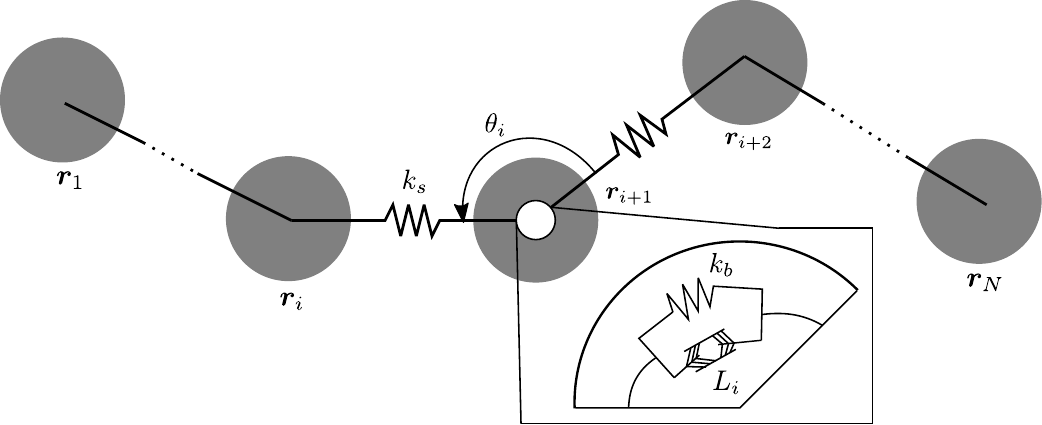}
    \caption{Schematic illustration of the bead-spring swimmer model.
    Beads of radius $a$ are connected with linear springs with spring constant $k_s$. The swimmer propels by the undulatory motion driven by the active torque $L_i$ assigned to the joints formed by three consecutive beads, acting together with a passive elastic bending energy term with bending stiffness $k_b$, which induces a force in a direction that tends to flatten the joints.}
\end{figure}

\subsection{Equations of motion}
The model microswimmer is suspended in a Newtonian fluid with viscosity $\eta$ and density $\rho$, and the Reynolds number $\mathrm{Re}=\ell v_0 \eta / \rho =(N-1)\rho \lmax/(\eta^2 \ell_0)$, with the characteristic bead velocity $v_0=\lmax/(\eta \ell_0^2)$,
is assumed to be very small, and hence we approximate the hydrodynamics by the Stokes flow limit.
We consider hydrodynamic interactions between different beads of the filaments in the far-field limit, given by the Oseen approximation of cross-mobilities $\bm{G}_{ij}$.
The instantaneous velocity of each bead is then computed as
\begin{equation}
  \bm{v}_i = \frac{\bm{f}_i}{6 \pi \eta a} + \sum_{j \neq i} \bm{G}_{ij} \bm{f}_j 
\end{equation}
where the first term for each bead $i$ considers self mobility, and the second term hydrodynamic interactions induced by the motion of all other beads, which create Stokeslets of the form
\begin{equation}
  \bm{G}_{ij} = \frac{1}{8 \pi \eta} \qty(\frac{\bm{I}}{r_{ij}} + \frac{\bm{r}_{ij} \otimes \bm{r}_{ij}}{r_{ij}^3}) 
\end{equation}
where $\bm{r}_{ij} = \bm{r}_i - \bm{r}_j$  and $r_{ij} = |\bm{r}_{ij}|$.
Note, while we allow the swimmer only to deform in 2D, we use the full 3D hydrodynamics.
The time evolution of the bead positions $\bm{r}_i$ is then obtained by solving
\begin{equation}
  \dv{\bm{r}_i}{t} = \bm{v}_i. \label{eq:motion}
\end{equation}

\subsection{Reinforcement learning} \label{sec:rl}
How the microswimmer deforms and eventually swims depends, in our case, only on the time sequence of active bending torques $L_i(t)$.
In this study, we use a reinforcement learning algorithm, PPO \cite{schulman2017proximal} (proximal policy optimization), to optimize the swimming strategy, i.e., the sequence of the $L_i(t)$.
The beads positions and velocities are used as the current state $\mathcal{S} = (\bm{r}_1, \bm{r}_2, \dots, \bm{r}_N, \bm{v}_1, \bm{v}_2, \dots, \bm{v}_N)$, while the active bending torques are the action $\mathcal{A} = (L_1, L_2, \dots, L_{N-2})$.

\rev{The applied torques on the joints of the microswimmer are not able to change instantaneously, but its joints are actuated}
by applying a new action only every time interval $T_a$, being referred to as the \textit{action interval}, and the action \rev{is continued} for this time until the next decision-making: only at times $t=0$, $T_a$, $2T_a$, $\dots$, a new action $\mathcal{A}$ \rev{is set} based on the current state $\mathcal{S}$ and the policy.
A net displacement in the $x$-direction is set as the immediate reward of this system as
\begin{equation}
  \mathcal{R}_t = (\bm{r}_g (t + T_a) - \bm{r}_g (t)) \cdot \bm{\mathrm{e}}_x 
\end{equation}
where $\bm{r}_g$ is the center of swimmer $\bm{r}_g = \sum_i^N \bm{r}_i/N$. 
The system is optimized based on trial-and-error to maximize the summation of the immediate reward
\begin{equation}
  \mathcal{S}_t = \mathcal{R}_t + \gamma \mathcal{R}_{t+1} + \gamma^2 \mathcal{R}_{t+2} + \cdots = \sum_{k=0}^{N_T} \gamma^k \mathcal{R}_{t+k}
\end{equation}
where $\gamma$  is the discount rate and $N_T$ is the number of iteration steps.
In this study, we used the following hyperparameters: learning rate as $3.0 \times 10^{-5}$, $16$ parallel environments, batch size as 16, epoch numbers as 20, and the discount rate $\gamma = 0.997$.

\subsection{Time Scales and Dimensionless parameters}
The governing equations are converted into a dimensionless form for the simulation.
The physical units are non-dimensionalized by using $\ell_0$ as the unit of length, $\lmax$ as the unit of torque (and energy), and $\eta$ as the unit of viscosity; hence, the unit of force is $\lmax/\ell_0$.
We thus choose as the unit of time the \textit{active viscous time scale} $t_0=\eta \ell_0^3/\lmax$,
which is the typical time a joint opens or closes by applying the active torque $\lmax$ to it in a viscous fluid of viscosity $\eta$.
The associated velocity scale, i.e.\ the typical speed the joint opens/closes, is then $v_0=\lmax/(\eta \ell_0^2)$.
Similarly, we define the \textit{passive viscous time scale} $t_p=\eta \ell_0^3/k_b$
as the typical time a joint closes by the presence of the passive restoring bending torque of strength $k_b$.
Finally, the third time scale is the action interval $T_a$.
In this work, we fix the dimensionless bead size  $a^* = a/\ell_0 = 0.1$ and spring constant $k_s^* = k_s \ell_0^2/\lmax = 10$ and focus on the effect of the following two dimensionless parameters, which can be written as the ratio of different time scales as
\begin{align}
  k_b^* &= \frac{t_0}{t_p} = \frac{k_b}{\lmax} \quad, \\
  T_a^* &= \frac{T_a}{t_0} = \frac{\lmax}{\eta \ell_0^3} T_a \, .
\end{align}
The first dimensionless parameter $k_b^*$ defines the relative stiffness of the swimmer's joint compared to the active torque.
Hence, for $k_b^\ast \ll 1 $, the passive restoring bending is negligible compared to the active bending,
whereas for $k_b^\ast \gg 1$ passive bending dominates and the active torque cannot induce any significant bending,
which is thus not a relevant regime in this study.
We set $k_b^* \ge 0.5$, i.e., a sufficiently stiff swimmer to avoid overlap of the beads, and thus the interesting range is of $\mathcal{O}(1)$, and we choose $k_b^\ast \in [0.5,2] $.
The second dimensionless parameter $T_a^*$ compares again two time scales, namely the action interval $T_a$ to the active viscous time scale $t_0$.
The \rev{movement of the swimmer can be controlled more precisely}
for smaller $T_a^\ast$,  since the dimensionless frequency of the decision-making $1/T_a^\ast$ is faster.
In particular, for $T_a^* \ll 1$
\rev{a new action is set} more or less immediately, whereas for $T_a^* \sim 1$ the action time is already comparable to the active joint deformation time.

The bead's positions are updated with a first-order Euler method by discretizing Eq.~\eqref{eq:motion} using time step $\Delta t = 10^{-3}t_0$.
The total simulation time is $\mathcal{T}=$ $10^3t_0$, corresponding to $10^6$ time steps.

\section{Results}
In the following, we first show the optimal strategy obtained using RL, and characterize the swimmer shape and torque actuation $L_i (t)$.
After characterizing the optimal swimming mode, we further analyze the swimming strategy by introducing deterministic torque actions into our simulation model.
We present the results for a swimmer of length $N=10$, for different parameters $k_b^*$ and $T_a^*$.
\rev{Note that the spring constant $k_s^* = 10$ is set sufficiently large to ensure that the swimmer is nearly inextensible, and the standard deviation of the relative body length modulations is for all cases $\leq 2\%$.}

\begin{figure}
    \centering
    \includegraphics[width=0.8\textwidth]{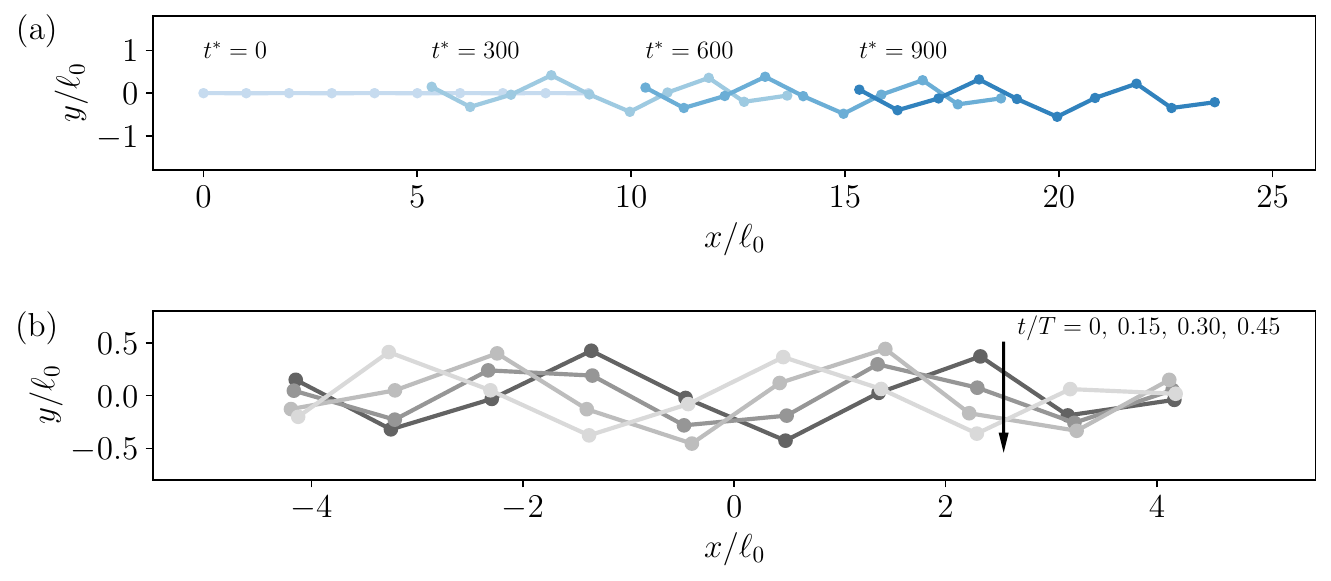}
    \caption{
      The optimized motion of microswimmer for (a) long term and (b) approximately half cycle from $t/T = 0$ (dark gray) to $ 0.45$ (lighter gray), under the dimensionless parameters $k_b^* = 0.5$ and $T_a^* = 0.1$. Note that the \rev{beating period} is $T^* = 2$ for this motion, and $t^*=900$ corresponds to a displacement after 450 beating cycles. \label{fig:swimming}
    }
\end{figure}

\subsection{Swimming optimization with reinforcement learning}
Using the reinforcement learning,
an optimized swimming strategy \rev{for the microswimmer is adapted}, identified by the actions $\mathcal{A}$, which depend on the parameters $k_b^*$, $T_a^*$, and the length $N$ of the swimmer.
As shown in Fig.~\ref{fig:swimming}(a), a microswimmer is initially at time $t^*=0$ aligned along the $x$ direction, and then propagates by optimized swimming gaits in the $+x$ direction.
As expected, the swimmer \rev{assumes} undulatory motion by performing \rev{time-}periodic shape changes of a certain period $T$, \rev{wavelength $\lambda$ and wave amplitude $b$, which depend on the parameters $k_b^*$ and $T_a^*$.
They are induced by a time-periodic actuation of the torques $L_i$, which assume bang-bang solutions, i.e.\ switching between $+\lmax$ and $-\lmax$ values, which are in nearly-constant phase lag between consecutive joints,
as detailed below.}

The undulatory wave pattern is shown in Fig.~\ref{fig:swimming}(b) by depicting four snapshots of the swimmer shape within a period $T$, \rev{which represent the time-periodic deformations, quantified by the angles $\theta_i(t)$ (not shown).}
Figure~\ref{fig:swimming}(a) shows the long-term displacement by repeating the undulatory motion for a few hundred cycles.
Note that the periods $T$ are determined from the Fourier transform of the actions $L_i(t)$.

\begin{figure}
  \centering
  \includegraphics[width=\columnwidth]{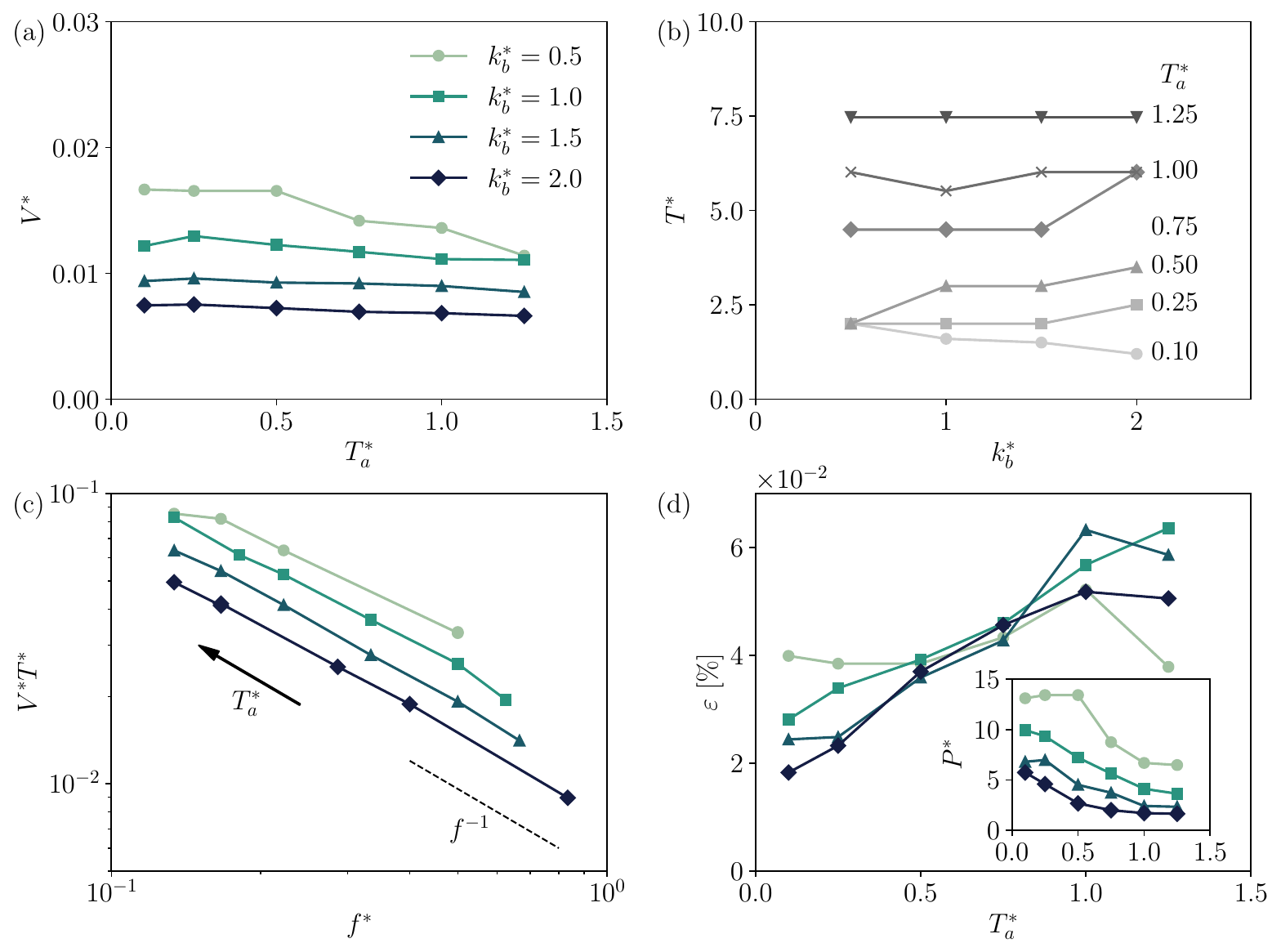}
  \caption{
    (a) Swimming velocities of the optimized swimming $V^*$.
    (b) Beating period $T^*$, which is the inverse of the beating frequency $f^* = 1/T^*$.
    (c) Gained distance per unit cycle $V^* T^*$ as a function of the beating frequency $f^*$. 
    (d) Swimming efficiency $\eta^*$. The inset figure shows the power expended by the swimmer $P^*$.
  \label{fig:velocity}}
\end{figure}

Since the RL algorithm maximizes the distance traveled in $+x$ direction in a certain time, this is equivalent to optimizing the center of mass velocities, calculated as $V = (\bm{r}_g (T) - \bm{r}_g (0))/T$.
The dimensionless velocities $V^* = V/v_0$ of our optimized swimmers for different bending stiffness $k_b^*$ and action interval $T_a^*$  are shown in Fig.~\ref{fig:velocity}(a).
The swimmer motions with varying $T_a^*$ but keeping $k_b^* = 1.0$ constant are also shown in Supplemental Movie S1.
We can clearly see that the swimmer can achieve a higher swimming speed for the smaller bending stiffness $k_b^*$, as discussed below.
Interestingly, the dependence of the speed on the applied action interval is relatively small, particularly for higher $k_b^*$.
For smaller $k_b^*$, however, the velocity slightly decreases with $T_a^*$.
This is somehow expected since a longer $T_a^*$ decreases the frequency of \rev{the adaptation of the action.}
Nevertheless, it first seems surprising that the velocity does not change significantly, despite the action interval being varied by a full order of magnitude.
Even more intriguingly, for a fixed $k_b^*$, the obtained optimized shapes and undulatory frequencies for the swimmers differ significantly despite the similar velocities.
This is first demonstrated in Fig.~\ref{fig:velocity}(b) where we show the dimensionless beating period $T^* = T / t_0$ is almost independent of $k_b^*$ but monotonically increases with the action interval $T_a^*$.
This indicates that 
\rev{a smaller action interval $T_a^\ast$ leads to shorter stroke cycles,}
while the corresponding swimming speeds are, in general, not altered significantly \rev{by changing $T_a^\ast$} (Fig.~\ref{fig:velocity}(a)).

To better understand the differences in the swimming strategy, we now break down the velocity into two components $V^* = V^*T^* \times f^*$, where $V^*T^*$ is the gained distance per unit cycle while $f^*=1/T^*$ is the beating frequency, as shown in Fig.~\ref{fig:velocity}(c): for a fixed $k_b^*$, the distance per cycle $V^*T^*$ is inversely proportional to the beating frequency $f^*$ obtained from the different action intervals.
The swimmers with large $f^*$ but small $V^* T^*$, which are the plots at the bottom right of the figure, have a strategy to beat many times though the propulsion per unit cycle is small.
On the other hand, swimmers with small $f^*$ but large $V^* T^*$ at the top left of the figure have a strategy to prioritize gain distance per cycle rather than the frequency.

While we optimize our microswimmers in terms of speed, we also evaluate the Lighthill efficiency $\varepsilon$, which is defined here as
\begin{equation}
  \varepsilon = \frac{6 \pi \eta N a V^2}{P}
\end{equation}
where $P$ is the total time-averaged power consumption \rev{of the microswimmer},
\begin{equation}
  P = \frac{1}{\mathcal{T}} \sum_i^N \int_0^\mathcal{T} \bm{v}_i \cdot \bm{f}_i \dd{t}
  \label{eq:P}
\end{equation}
\rev{which coincides with the total drag-associated dissipated energy in the system,}
where $6 \pi \eta N a$ is approximately the Stokes friction coefficient of a passive $N$ bead microswimmer \cite{Hartl2025}.
\rev{Note that the passive spring and bending forces contributing to the bead forces $\bm{f}_i$ do not produce any net work done by the microswimmer over a period, but only the active bending forces.}
Figure~\ref{fig:velocity}(d) demonstrates that the efficiency $\varepsilon$ shows a generally increasing trend with the action interval.
The efficiency increases for larger $T_a^*$ since the power consumption decreases with $T_a^*$, as shown in the inset of Fig.~\ref{fig:velocity}(d), while the velocities are almost independent of $T_a^*$.
This suggests that the swimming strategies found for small $T_a^*$ are recommended for (potentially) faster swimming, while the strategy with large $T_a^*$ is recommended for more efficient swimming.
Figure~\ref{fig:velocity}(d) shows that the efficiency is in order of $10^{-2}$\%, and the efficiency is nearly the same with the ciliate swimmers, which have an efficiency $10^{-2}-10^{-1}$\% \cite{omori2020swimming}.
However, the value is smaller than that of \textit{E. Coli} and sperms \cite{omori2020swimming} or Najafi-Golestanian three-sphere swimmer \cite{nasouri2019efficiency} that have efficiency in an order $\sim$1\%.
\rev{The main reason for the relatively small efficiencies is the relatively large distance between the spheres which limit the hydrodynamic coupling and hence the effective generated thrust. Furthermore in our approach we optimize for swimming speed and not for efficiency.}

\begin{figure}
  \centering
  \includegraphics[width=\textwidth]{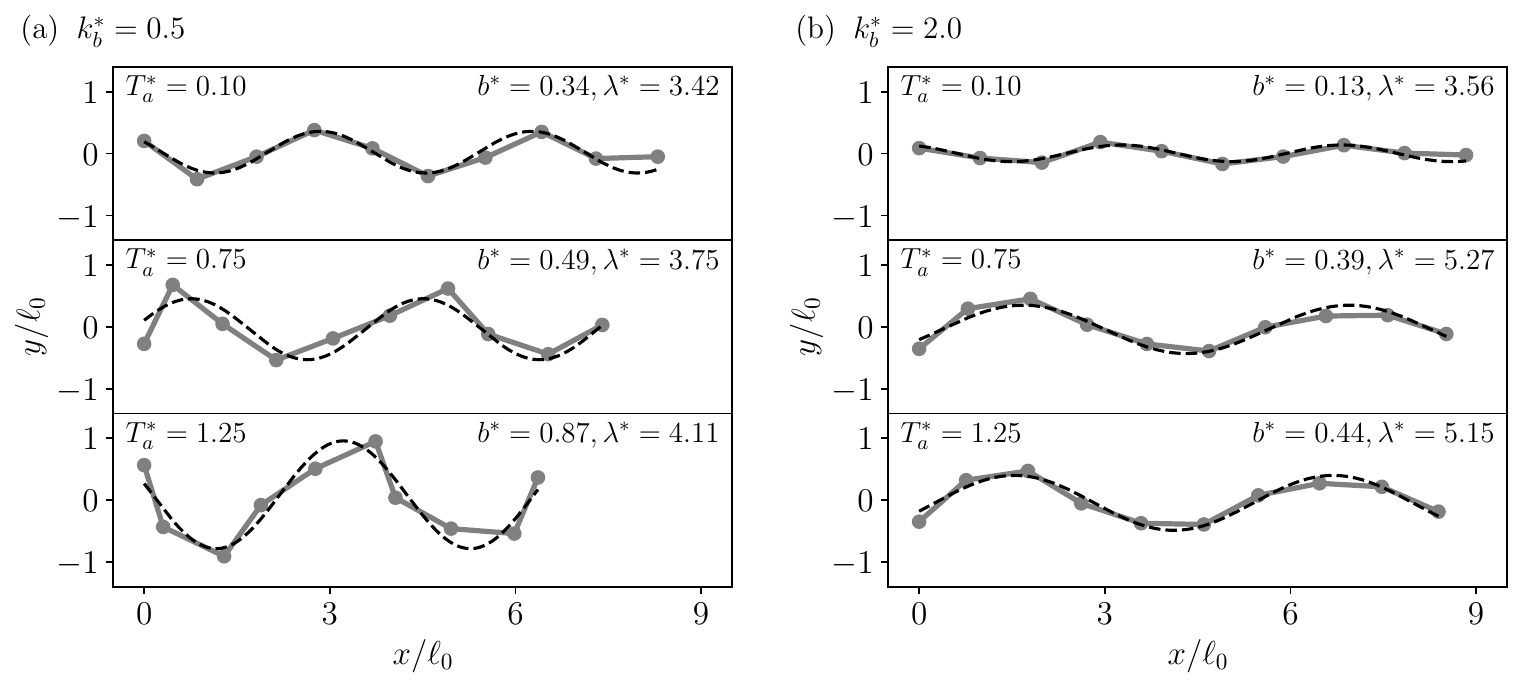}
  \caption{Swimmer shape for (a) $k_b^* = 0.5$ and (b) $k_b^*=2.0$. Gray connected beads show the swimmer shape, while black dashed lines show the fitting results. The instantaneous amplitude $b^*$ and wavelength $\lambda^*$ are shown in the upper right corner of each figure. \label{fig:swimmer_shape}}
\end{figure}
\begin{figure}
  \centering
  \includegraphics[width=\textwidth]{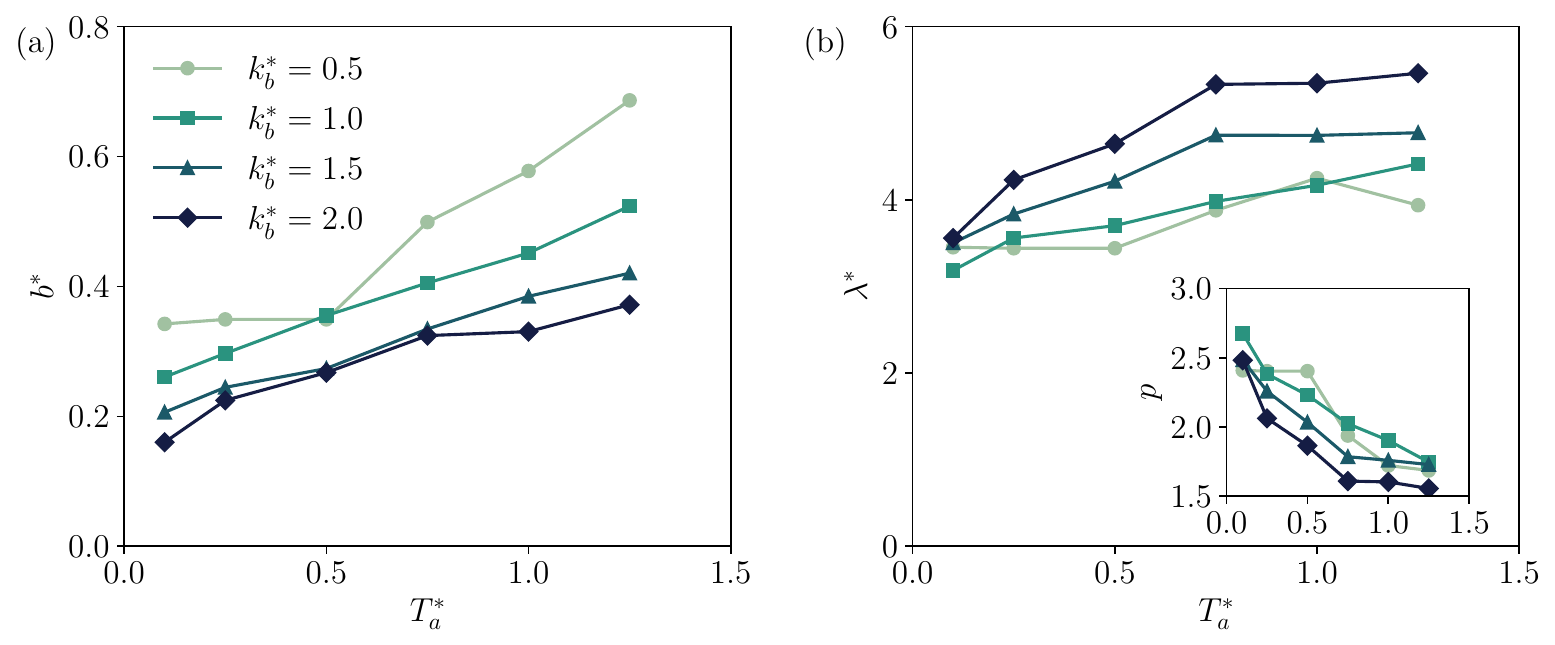}
  \caption{Geometric features of the optimized swimming mode: (a) amplitude $b^*$ and (b) wavelength $\lambda^*$. The inset figure shows the number of waves in the swimmer's body. \label{fig:shape}}
\end{figure}

To get a more complete picture, we characterize the shapes of the different swimmers, depending on the parameters.
Figure~\ref{fig:swimmer_shape} shows the swimmer deformation for (a) $k_b^*=0.5$ and for (b) $k_b^*=2.0$, each for three different applied action intervals $T_a^*$.
The shapes of the swimmers can generally be fitted with a single dimensionless wavelength $\lambda^*$ and amplitude $b^*$.
We use sinusoidal fits applied to the instantaneous bead positions,
\begin{equation}
  y(x) = b \sin(kx + d) = b \sin(\frac{2 \pi x}{\lambda} + d)
\end{equation}
where $k = 2 \pi/ \lambda$ is the angular wave number and $d$ is the phase.
This fitting function is well-suited for smaller amplitudes and approximately still works for larger amplitudes where the shapes can be more saw-teeth-like and show some end effects.
The geometric features of the swimmers are summarized in Fig.~\ref{fig:shape}.
\rev{The amplitude $b^*$ and the wavelength $\lambda^*$ increase with the action interval $T_a^*$, while the number of waves along the swimmer length $p = \ell_\mathrm{all}/\lambda$ decreases, where $\ell_\mathrm{all}$ is the $x$-directional head-to-tail distance.}
Two geometric values have opposite trends by increasing $k_b^*$: the amplitude $b^*$ decreases while $\lambda^*$ increases.
This trend is intuitive to understand since large bending stiffness would lead to smaller bending that leads to smaller amplitude and larger wavelength.

We can now approximate the swimmer shape by a sinusoidal traveling wave, characterized by amplitude $b$, angular wave number $k$, and angular frequency $\omega = 2\pi /T$, which are obtained as fit parameters from the actual dynamic swimmer shapes.
For the fitted shape, the swimming velocity can be calculated as $V=V(b,k,\omega) = \omega F(b,k)$, which scales linearly with $\omega$ and depends on a shape-dependent function $F(b,k)$, which for a continuum wave increase with $b$ and $k$ for sufficiently small amplitude.
For example, in the Taylor sheet approximation \cite{Taylor1951} for an infinitely long waving sheet, the leading order swimming speed is given by $V_\mathrm{TS} = k \omega b^2/2$.
For one-dimensional waving filaments, the dependence on $k$, $b$ and $\omega$ is the same as for the Taylor sheet; however, with a prefactor $\kappa <1$, which depends on the specific model, such as filament thickness for continuum lines, or on $a^\ast$, as in our case.
This formula may be modified for large amplitudes and for finite microswimmer length, as also discussed in previous literature \cite{Lauga2009a}.

\begin{figure}
    \includegraphics[width=\textwidth]{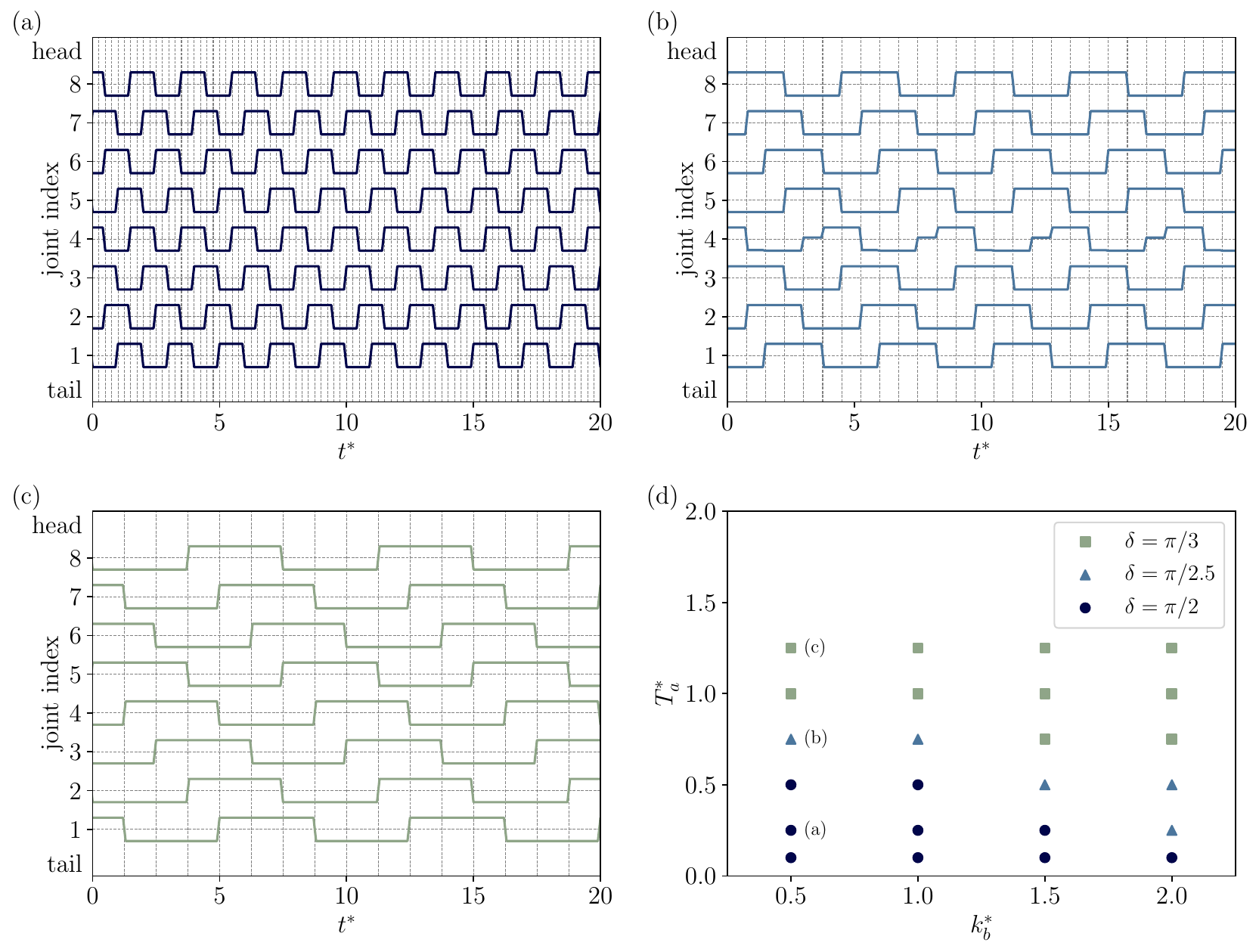}
    \caption{
    (a)-(c) Time histories of the torque input for different conditions. The horizontal axis shows the elapsed time, while the vertical axis shows the torque of 8 joints. Except for the 3rd joint in (b), the torques always have either the maximum $L_i/\lmax = +1$ or the minimum $L_i/\lmax = -1$ values. The vertical dotted lines in the figures show the timing that the action can be switched.
    (d) Phase diagram of the phase differences with the neighboring joint.  
    \label{fig:torque}}
\end{figure}

To summarize, the swimming strategies with small $T_a^*$ beat faster with small deformation, while the strategy with large $T_a^*$ assumes large deformation though the beating is slow. 
The outcome of the dynamic swimmer shape and, consequently, its swimming speed depends on the policies obtained by the RL, quantified by the torque actuation $L_i(t)$.
Fig.~\ref{fig:torque}(a)-(c) show the obtained time evolution of the torque actuation $L_i(t)$ that is applied to each bead $i=1,\dots,N-2$ for different $k_b^*$ and $T_a^*$. 
It can be seen that the values almost always flip between either the maximum $L_i^* = L_i/ \lmax = +1$ or minimum $L_i^* = -1$ value and that the torque patterns are propagated from the head to tail direction with nearly constant phase differences $\delta$.
In the case of Fig~\ref{fig:torque}(a), for instance, the phase difference $\delta$ between the neighboring joints can be estimated as $\delta = \pi/2$ since $i$-th and $(i+4)$-th joints display the same pattern; in other words, every four neighboring joints form a single wave. 
Interestingly, the phase difference $\delta$ in general depends on $k_b^*$ and $T_a^*$ as summarized in Fig.~\ref{fig:torque}(d). 
For relatively small $T_a^*$, which favor high frequency small amplitude oscillations as discussed before, the phase difference is $\delta = \pi/2$.
By increasing the action interval $T_a^*$, for a fixed $k_b^*$ the phase difference suddenly jumps from $\pi/2$ to $\pi/2.5$ ($=2\pi/5$), and eventually to $\pi/3$.
The discontinuous jump in the phase difference can be explained by the way that the action is actually chosen: each joint can change the torque input at every period $T_a^*$, and the timing of the switching action is synchronized as an internal clock.
Therefore, the resultant phase differences have to be $2 \pi/M$, where $M$ is an integer number.
If the swimmer can change the action at any time, the transition in the phase difference would be continuous.

\begin{figure}
  \includegraphics[width=\textwidth]{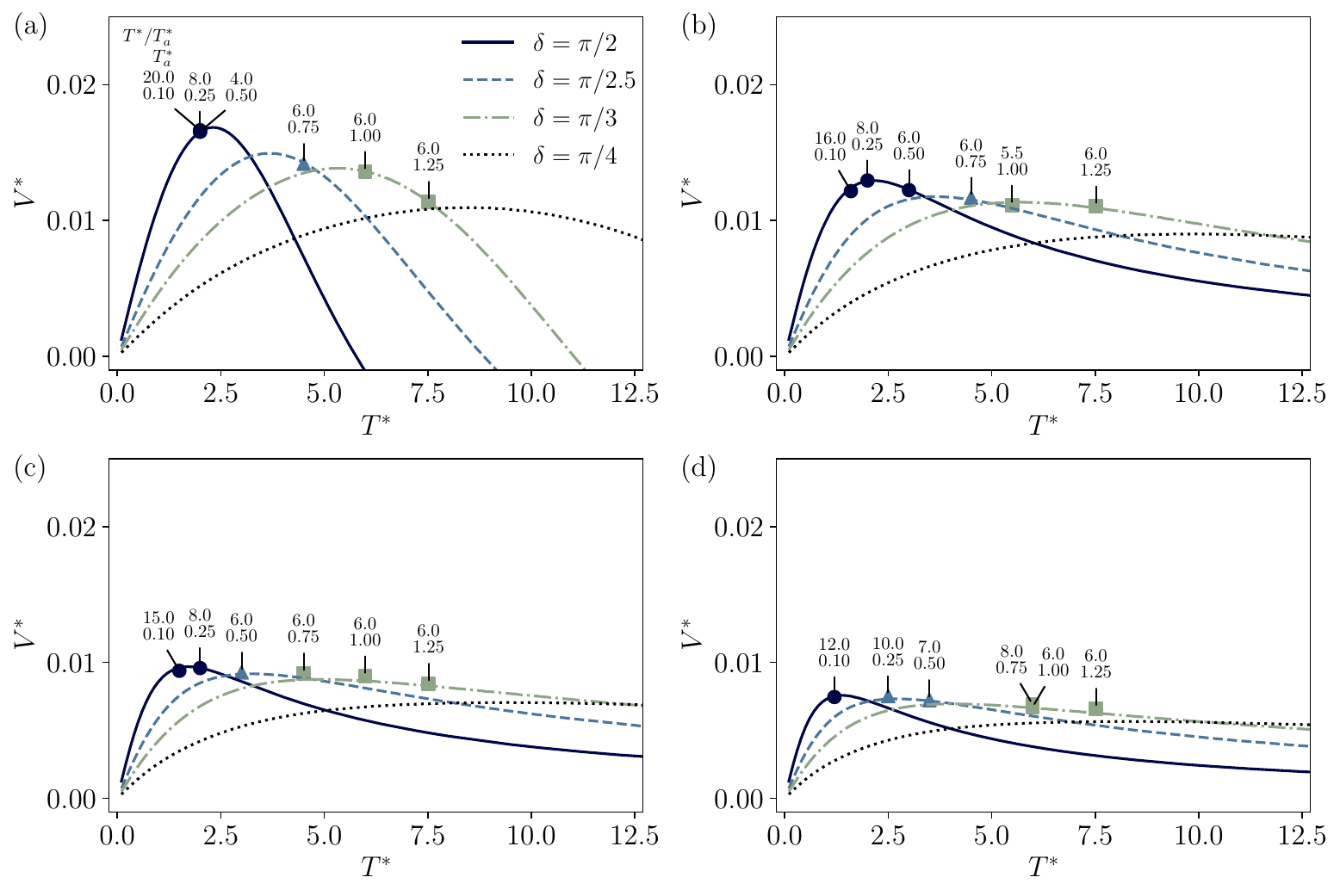}
  \caption{Swimming velocity $V^*$ of the model with different dimensionless bending stiffness (a) $k_b^* = 0.5$, (b) $1.0$, (c) $1.5$ and $2.0$. The points represent the result of the reinforcement learning, while the lines show the simulation results with given torque inputs. The symbols represent the phase differences of the torque patterns: $\pi/2$ (\textbullet), $\pi/2.5$ ($\blacktriangle$) and $\pi/3$ ($\blacksquare$). \rev{The annotated numbers with the leader lines represent the conditions set in RL; the number of actions per period $T^*/T_a^*$ (top) and the action interval $T_a^*$ (bottom).} \label{fig:comparison}}
\end{figure}

Hence, because of the finite action interval, the \rev{beating period} $T$ can now be described by the integer multiple of the action interval, $T = n_a T_a$, where $n_a$ is the number of actions per unit cycle. 
For instance, $n_a = 8$ for Fig.~\ref{fig:torque}(a) and $n_a = 6$ for Fig.~\ref{fig:torque}(b)-(c).
As also shown later in Fig.~\ref{fig:comparison}, the optimized swimmers prefer even numbers of $n_a$ with $n_a \geq 4$.
To understand this, let us assume that each joint will pick up the active torque either $+\lmax$ and $-\lmax$ for each action interval $T_a$, for simplicity.
To swim straight, a single joint should have the same number of $+\lmax$ and $-\lmax$ actions during a single cycle; i.e., the number $n_a$ should be even.
For $n_a=2$, a single joint will have actions $(+\lmax, -\lmax)$ during a single cycle $T = 2T_a$. 
In this case, the swimmer cannot produce the propagating wave since there are only two patterns for the actions: in-phase ($+L_\mathrm{max}$, $-L_\mathrm{max}$, $+L_\mathrm{max}$, $-L_\mathrm{max}$, $\cdots$) or opposite phase ($-L_\mathrm{max}$, $+L_\mathrm{max}$, $-L_\mathrm{max}$, $+L_\mathrm{max}$, $\cdots$).
In order to achieve the propagating wave-like deformation and the net propulsion, at least $n_a = 4$ is required.
For $n_a=4$, the propulsion is achieved because of four different patterns for the actions: ($+\lmax, +\lmax, -\lmax, -\lmax, \cdots$), ($-\lmax, +\lmax, +\lmax, -\lmax, \cdots$), ($-\lmax, -\lmax, +\lmax, +\lmax, \cdots$) and ($+\lmax, -\lmax, -\lmax, +\lmax, \cdots$).
These four patterns are imposed on each successive four particles, resulting in $\pi/2$ phase differences with the neighboring particles.
For $n_a=6$, the phase difference will be $\pi/3$.

\subsection{Swimming with different phase differences}
Although we can understand now the basic geometric features of the optimized swimming, it is still unclear why for a given $k_b^*$ and $T_a^*$ a specific value of $\delta$ is picked, as shown in Fig.~\ref{fig:torque}(d).
Certainly, it has to do with the hydrodynamic interactions between the beads, which are responsible for the obtained optimized swimming speed.

Motivated by the simple square-like wave patterns of the actions $L_i(t)$ found by the RL, as shown in the previous sections, we now run simulations with simple predefined square-like wave pattern policies $L_i(t)$ of the form
\begin{equation}
  L_i^* (t) = \mathrm{sgn} \left( \sin (\frac{2 \pi t^*}{T^*} + i \delta) \right).
  \label{eq:Lsgn}
\end{equation}
for different phase difference $\delta$ and beating period $T^*$, for the different bending stiffness $k_b^*$.
Figure~\ref{fig:comparison} shows the comparison between the swimming velocity $V^*$ of the reinforcement learning and that from the simulations \rev{using the torque inputs from Eq.~(\ref{eq:Lsgn})}.
The simulation result shows that lines with different phase differences reach the maximum velocity at a certain wave period, which we refer to as $T_\mathrm{optimal}$.
We show curves for four different phase differences $\delta$, demonstrating that the values of  $T_\mathrm{optimal}$ increase with decreasing $\delta$. 
The phase difference $\delta=\pi/2$ corresponds to the maximum velocity since it maximizes the time symmetry breaking; e.g.\ in contrast, there is no propulsion for in-phase ($\delta = 0$) and anti-phase ($\delta=\pi$) since here time symmetry is not broken.
Note that the optimum velocity decreases for phase differences $\delta > \pi/2$ (data not shown).

As also shown in Fig.~\ref{fig:velocity}(c), there is a trade-off relationship between the beating frequency $f$ and the gained distance per cycle $VT$, and $T_\mathrm{optimal}$ emerges due to this relationship: since the increase in $f$ would decrease $VT$, the resultant velocity $V = VT \times f$ should have the maximum value at the certain frequency.
We can now estimate the optimal period $T_\mathrm{optimal}$ from the relaxation time of the deformation.
We first consider a single joint consisting of three initially aligned beads (passive equilibrium state with bending angle $\theta_0=\pi$, see Appendix A), which is bent now by a constant active torque $L^\ast=+1$. Then the angle $\theta(t)$ evolves as
\begin{equation}
  \theta(t) = \pi - \frac{1}{k_b^*} + \frac{1}{k_b^*} \exp \qty(- \frac{k_b^*}{\pi a^*} t^*). \label{eq:theta}
\end{equation}
in the limit of small angle deviations $\phi(t) \equiv \theta_0 - \theta(t) \ll 1$ (otherwise $\theta (t)$ can be solved numerically, constant bead-bead distances and when hydrodynamic interactions are ignored.
Hence, the angle $\theta$ converges to $\theta_f = (\pi - 1/k_b^*)$ exponentially with relaxation time $\tau^* = \pi a^*/k_b^*$. As expected, larger bending stiffness $k_b^\ast$ leads to smaller deformations and smaller relaxation times within our approximations. This is in line with the results shown in Fig.~\ref{fig:comparison}(a)-(d), where the optimal period $T_\mathrm{optimal}$ is smaller for larger $k_b^*$.
Now, when we, for example, consider a model system of four beads and two joints, the relaxation time of the bending $\tau^* = 3 \pi a^*/k_b^*$ is now longer compared to the relaxation of a single joint, as derived in Appendix A.
In this case, two adjacent beads are subjected to torques of the same sign, and the bending forces are partially counteracted and hindered, resulting in an extended relaxation time.
Altogether, the swimmer takes a longer time to be bent for deformation with a larger wavelength $\lambda$, and this contributes to a longer $T_\mathrm{optimal}$ for smaller $\delta$.
This is in agreement with previous studies showing that the bending relaxation time for an elastic rod grows with its wavelength
\cite{Wiggins1998,munk2006dynamics}.
As a consequence, further increasing the action interval is expected to lead to even longer optimum wavelengths.

\begin{figure}
  \centering
  \includegraphics[width=0.85\textwidth]{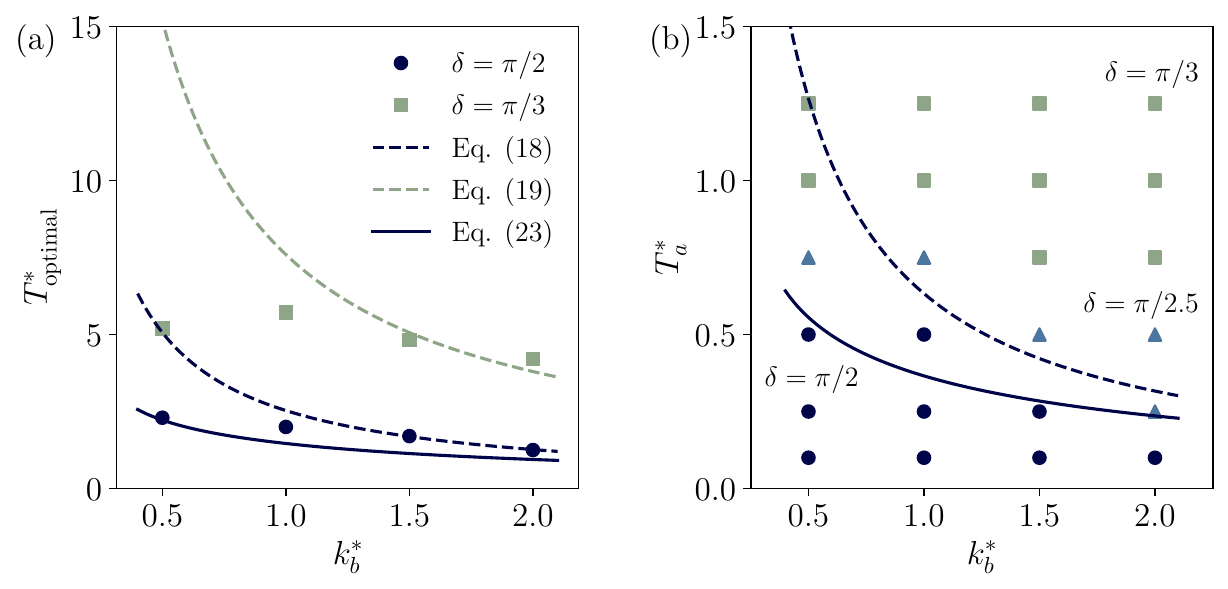}
  \caption{(a) The optimal period $T_\mathrm{optimal}^*$ for different bending stiffness $k_b^*$. Theoretical predictions that are shown with lines agree well, especially at large $k_b^*$, with the simulation results shown with points. (b) Phase diagram of the phase differences with the theoretical predictions that divide the two phase differences, $\delta = \pi/2$ and $\delta = \pi/2.5$.\label{fig:prediction}}
\end{figure}

Now we will use the estimated relaxation time $\tau^\ast$ to relate it to the optimal period $T_\mathrm{optimal}$.
For efficient swimming using a cyclic bending process, it is necessary not to wait until the joint reaches its equilibrium angle $\theta_f$, which would take an infinitely long time, before reversing the torque.
Instead, the torque has to be reversed at some point before reaching $\theta_f$ at a finite time to bend the joint in the opposite direction.
At a time $t_\alpha$, an initially flat joint has bent to a fraction $1-\alpha$ of its final value, $\alpha= \exp(-t_\alpha/\tau^*) $, hence occuring at a time $t_\alpha=\tau^\ast \ln (\alpha^{-1})$.
To reach the optimum period $T^*_\mathrm{optimal}$, \rev{for the following we assume that a certain $\alpha$ exists. The specific value of $\alpha$ is discussed below.}
Then, since this single bending process can be regarded as the quarter of a cycle, $T^*_\mathrm{optimal}$ can be evaluated as
\begin{equation}
  T^*_\mathrm{optimal} = 4 t_\alpha = 4 \tau^* \ln (\alpha^{-1}) = \frac{4 \pi a^*}{k_b^*} \ln(\alpha^{-1}). \label{eq:t_optimal1}
\end{equation}
Similarly, the optimal period that can be predicted from the two-joint (four beads) structure can be evaluated as
\begin{equation}
  T^*_\mathrm{optimal} = \frac{12 \pi a^*}{k_b^*} \ln(\alpha^{-1}). \label{eq:t_optimal2}
\end{equation}
Figure \ref{fig:prediction}(a) compares the optimal periods that are obtained from the numerical result with the theoretical predictions, Eqs.~\eqref{eq:t_optimal1} and \eqref{eq:t_optimal2} with $\alpha = 2/15$. 
The figure shows that the theory has good agreement, especially for small deformations with large $k_b^*$: the optimal period for the phase differences $\delta = \pi/2$ and $\delta = \pi/3$ can be roughly estimated by the relaxation time of bending of three and \rev{four beads} structures, respectively. 
The prediction fails for large deformations since we assumed the small bending $\phi \ll 1$ during the derivation.

The full picture of the swimmer strategy can be summarized as follows.
Since the swimming strategy with $\delta = \pi/2$ gives the global maximum velocity as shown in Fig.~\ref{fig:comparison}, 
this strategy with $T^* = T_\mathrm{optimal}^*$ \rev{would be picked up} if the dimensionless action interval $T_a^* = \lmax T_a/(\eta \ell_0^3)$ is small enough; i.e., 
this strategy \rev{is preferred} when the action interval $T_a$ is small compared \rev{to} the relaxation time of the system.
For conditions with large $T_a^*$, the strategy with $\delta=\pi/2$ is no longer fast since the bending process is so fast that the swimmer wastes time waiting even after the deformation has already been saturated. 
Instead, inputs with smaller $\delta$ \rev{are a better choice now}, which results in longer wavelengths, since the relaxation time is longer and continuous propulsion can be achieved.
As a result, the swimmer discretely changes its phase difference from $\delta=\pi/2$ to smaller values as shown in Fig~\ref{fig:comparison}, and hops across the peaks of different phase differences. 

\begin{figure}
  \centering
  \includegraphics[width=0.75\textwidth]{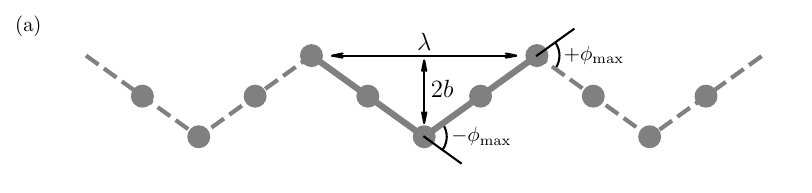}
  \includegraphics[width=0.90\textwidth]{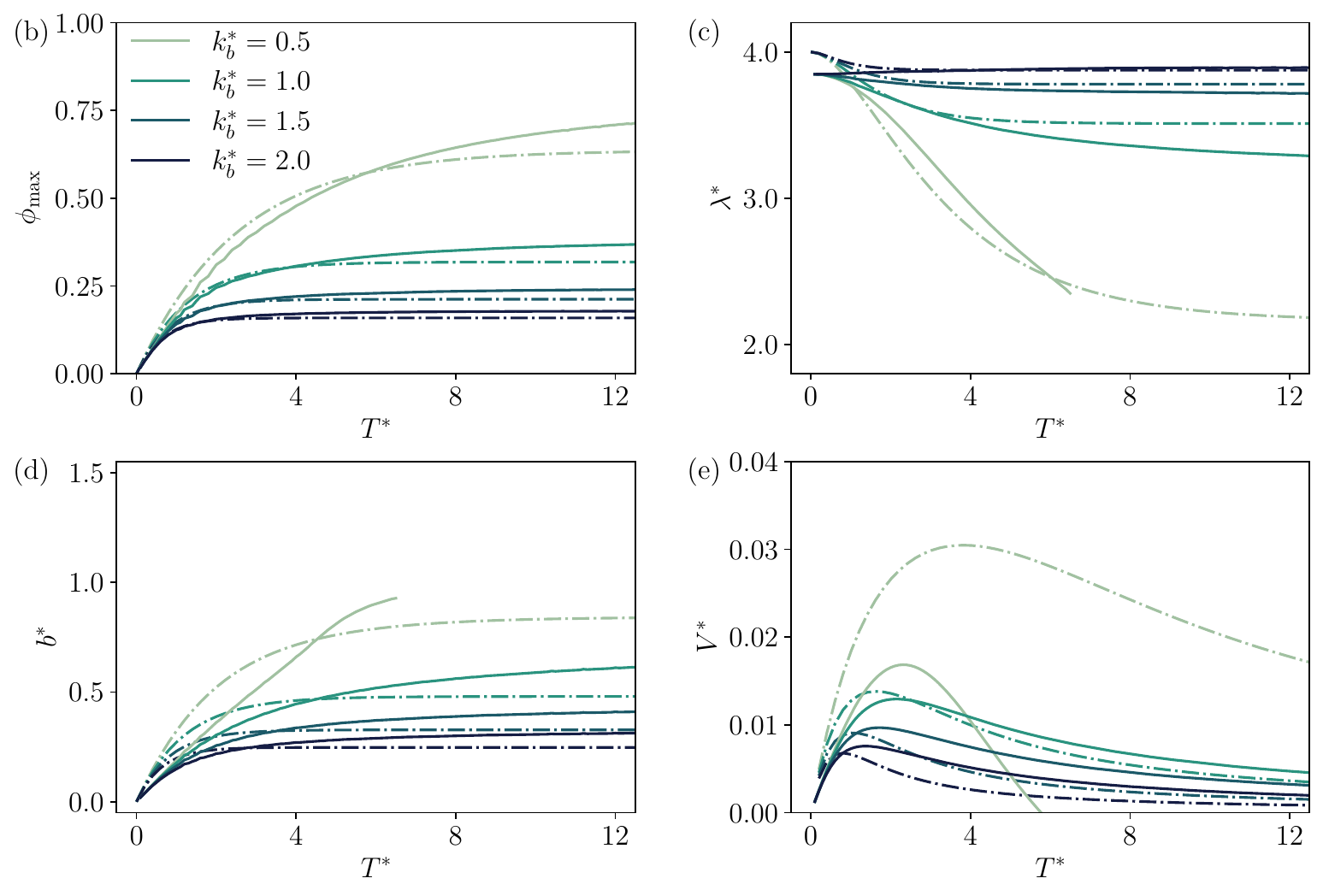}
  \caption{(a) A schematic showing the swimmer deformation under a phase difference $\delta = \pi/2$. Comparison of the swimmer deformations of theory \rev{(Eqs.~\eqref{eq:lambda}-\eqref{eq:velocity}, dash-dotted line) and simulations (using Eq.~\eqref{eq:Lsgn}, solid line)}: (b) maximum value of the joint angle $\pmax$, (c) wavelength $\lambda$, (d) amplitude $b$, and (e) velocity $V$. Note that the line ends around $T^* = 5$ for $k_b^* = 0.5$ for (c) and (d) since the swimmer shape cannot be fitted with a simple sinusoidal function. \label{fig:toymodel}}
\end{figure}

\subsection{Estimation on the optimal beating period}
Up to now, we have obtained estimates for the optimum behavior, but we cannot estimate the prefactor related to $\alpha$.
Clearly, finding $\alpha$ is related to the full hydrodynamic problem since hydrodynamic interactions alone determine the swimming speed.
We now construct a simple model to roughly estimate the period $T_\mathrm{optimal}$ for the important case of phase difference $\delta=\pi/2$.
For small amplitudes, the swimming speed can be written as $V^\ast = C V^\ast_\mathrm{TS} =C k^\ast \omega^\ast (b^\ast)^2/2 =2 \pi^2 (b^\ast)^2/(T^\ast \lambda^\ast)$ where $V^\ast_\mathrm{TS}$ is the famous Taylor-sheet swimming velocity, and $C<1$ is a prefactor capturing the effective thrust; for example when a continous curve is given, in the resistive-force-theory approximation $C$ depends on the coefficient of parallel to perpendicular drag coefficient of a slender rod.
In our case, it essentially depends on $a^\ast$.
To estimate $V^\ast$, we first build equations how the geometric parameters change with the \rev{beating period} $T^\ast$, $b(T^\ast)$, and $\lambda(T^\ast)$.
When the phase difference is $\delta=\pi/2$, the swimmers have the saw-like shape as shown in Fig.~\ref{fig:swimmer_shape} due to the phase difference, and this saw-like shape can be approximately estimated via the following simplifications.
When $i$-th joint angle is $\phi_i = +\pmax$, $(i+2)$-th and $(i+4)$-th joint is expected to have angles $\phi_{i+2} = -\pmax$ and $\phi_{i+2} = +\pmax$ as shown in Fig.~\ref{fig:toymodel}(a) since these joints are in opposite phase and in-phase, respectively. 
Thanks to \eqref{eq:phi}, the maximum values of the joint angle $\phi = \pi - \theta$ during the deformation can be estimated under $\phi \ll 1$ as
\begin{equation}
  \pmax (T) = \frac{1}{k_b^*} - \frac{1}{k_b^*} \exp \qty(-\frac{k_b^*}{4 \pi a^*} T^*). \label{eq:phi_max}
\end{equation}
Note that we substitute $t = T/4$ to \eqref{eq:phi} since the joint angle is estimated to reach from $\phi = 0$ to $\phi = \pmax$ with a period quarter. 
Since $(i+1)$-th and $(i+3)$-th joint angles are expected to be $\phi_{i+1}=0$ due to $\delta=\pi/2$, four successive neighboring joints would have angles $(\cdots, +\pmax, 0, -\pmax, 0, \cdots)$, and these joint angles would result in the saw-like shape as shown in Fig.~\ref{fig:toymodel}(a). 
Now, the wavelength and the amplitude can be described as the function of $\pmax$ from a simple geometry as
\begin{align}
  \lambda (\phi_\mathrm{max}) &= 4 \ell_0 \cos \frac{\phi_\mathrm{max}}{2}, \label{eq:lambda} \\ 
  b (\phi_\mathrm{max}) &= \ell_0 \sin \frac{\phi_\mathrm{max}}{2}.
\end{align}
Figure~\ref{fig:toymodel}(b) compares the joint angle $\pmax$ of the theory \eqref{eq:phi_max} and the simulations, and they have reasonably good agreement, especially when the deformation is small.
Note that the angle of the simulation is calculated by averaging the maximum angle of the time history of all joints $\phi_i (t)$.
Using the max angle $\pmax$, the wavelength $\lambda^*$ and the amplitude $b^*$ can also be well-predicted as shown in Fig.~\ref{fig:toymodel}(c) and (d). 
Since we have the geometric and temporal information of the swimmer's motion, we can now simply estimate the velocity from the Taylor swimming sheet $V_\mathrm{TS} = 2 \pi^2 b^2/(T \lambda)$ as 
\begin{equation}
  V^* = C
  \frac{\sin^2 (\frac{1}{2 k_b^*} (1 - \exp(- k_b^* T^* /(4 \pi a^*))))}
       {T^* \cos (\frac{1}{2 k_b^*} (1 - \exp(- k_b^* T^* /(4 \pi a^*))))} \label{eq:velocity}
\end{equation}
where $C$ is the additional prefactor to correct the difference between the swimming sheet and the bead-spring swimmer.
Figure~\ref{fig:toymodel}(e) compares the simulation result with Eq.~\eqref{eq:velocity} with \rev{$C=1/6$}.
Though there are no quantitative agreements, the equation well captures the profile that initially increases and gradually decreases with $T^*$, and the optimal period $T_\mathrm{optimal}$ is well captured.
The optimal period $T_\mathrm{optimal}^*$ can now be estimated by numerically solving \rev{$\dv*{V^*}{T^*} = 0$}, and Fig.~\ref{fig:prediction}(a) shows that the theoretical model has good agreement with the numerical results. 
Finally, Fig.~\ref{fig:prediction}(b) shows the phase diagram of the selected phase difference $\delta$, which is the same as Fig.~\ref{fig:torque}(d), with the theoretical predictions.
The figure shows that the predictions well capture the transition in the phase difference between $\delta = \pi/2$ and $\pi/2.5$.

\section{Discussion and Conclusion}

We have introduced a model that optimizes the dynamic shapes of planar beating microswimmers using reinforcement learning.
In contrast to previous optimization and RL approaches, we introduced (i) a local constraint on the maximally allowed \rev{torque} input accessible for filament deformation, characterized by the maximum torque strength $\lmax$.
Second, (ii) the locally applied torques $L_i$ \rev{are the actions taken for a given filament} configuration, \rev{and not pre-defined quantities of a certain periodicity.}
Third (iii), in order to capture biologically relevant potential delay between reception of the internal state and the swimmer's action, we investigated the effect of a so-called action interval which only allows the filament to update its action, i.e.\ the active bending-generated torques $L_i$, between certain time intervals $T_a$.
Fourth, (iv) we thus do not set a beating frequency $\omega$ of the filament from outside, but the periodic beating of the filament and the associated frequency $\omega$ is an emerging property from \rev{the} decisions on the actions $L_i$, as well as the overall shape, i.e.\ the beating amplitude $b$ and wavelength $\lambda$.
We demonstrate that the beating strategy $(\{\omega, b, \lambda \})$ under the constraint of maximum applied torques and a finite action interval depends strongly on the bending stiffness $k_b$ of the filament, as well as on the action interval $T_a$.
We show that the diverse beating strategies are a result of three competing time scales in the system (active viscous bending time, passive viscous bending time, and action interval), quantified by two new dimensionless parameters, the passive/active bending ratio $k_b^\ast$, and the ratio of action interval and active viscous bending time $T_a^\ast$.
As visualized in Fig.~\ref{fig:swimmer_shape} and quantified in Figs.~\ref{fig:velocity} and \ref{fig:shape}, the optimal strategies are determined by a non-trivial combination of both dimensionless parameters.
The optimal strategies \rev{found by RL are for all cases periodic bang-bang torque solutions, i.e.\ alternating sequences} of maximum torques $+\lmax$ and $-\lmax$ to the different joints \rev{operating at a specific optimal periodicity $T$, and} which differ by a specific phase lag $\delta$.
Using a simple model based on naive swimmer policies, we demonstrate that for different phase lags, an optimum period of applied actions exists, where the global maximum always occurs for the phase lag $\delta = \pi/2$.
Indeed, this phase lag is the preferred one obtained by RL, as long as it is compliant with the action interval, which we show has to be at least four times smaller than the period of the applied actions to result in a traveling wave solution.
However, we show that for increasingly large action intervals, the phase difference $\delta$ decreases.
This arises because the deformation modes with smaller phase lags exhibit longer relaxation times, leading swimmers with larger action intervals to preferentially adopt these longer wavelength modes and thus optimize the use of their limited beating time.
In contrast \rev{to previous work on optimizing undulatory swimming strokes, in our framework} frequency and wavelength are emerging quantities from the RL optimization, and not preset, and our results are a consequence of a combination of the two dimensionless parameters $k_b^\ast$ and $T_a^\ast$
\rev{by assuming a local constraint on the torque inputs.}

Our work also provides new insights into the optimal swimming behavior of undulatory microswimmers in fluids of varying viscosity $\eta$, while maintaining constant bending stiffness and action interval.
First, the fact that both the active and passive bending timescales depend on viscosity similarly leads us to conclude that, under the constraint of a maximum active torque independent of viscosity, the optimum swimming strategy, and hence the swimmer's shape and frequency, do not change with viscosity.
However, we observe that the competition of action interval to viscosity-dependent bending time scale plays a crucial \rev{role}.
This offers a new way for shape control, in contrast to current visco-elastic coupling models of undulatory beating, see e.g.\ Ref.~\cite{Pierce2025,Taketoshi2025}.
In our case, the increase of viscosity leads to an increase in the beating frequency, and a decrease of the beating amplitude and wavelength, because $T_a^\ast \sim \eta^{-1}$. 

\rev{Finally we want to note, while we present results for a swimmer length of $N=10$, we confirmed by using $N$ between $4$ and $15$ in additional RL runs that the general strategy remains the same, i.e.\ phase-shifted bang-bang solutions with $T_a$- and $k_b$-dependent optimum periodicity and phase lag which converge for increasing $N$.}

Altogether, we have introduced a new framework to model the optimization of undulatory swimming based on constraints on the local maximum torque input.
\rev{For the future, it would be interesting to relate and extend our analysis to waving filaments with hydrogels \cite{boiardi2024minimal}, colloidal chains \cite{yang2020reconfigurable}, DNA \cite{dreyfus2005microscopic} or DNA origami \cite{suzuki2020large}, and magnetic soft robots \cite{pramanik2024nature}.
It would also be valuable to combine this with existing active-bending models of undulatory swimming \cite{Pierce2025} and to extend the modeling of hydrodynamic interactions using resistive-force theory or regularized Stokeslets.} 

\section*{Acknowledgement}
This work is supported by Japan Society for the Promotion of Science KAKENHI Grants (21H05879, 22H01402, 23H04418, 23K26040) and Japan Science and Technology Agency PRESTO Grant (JP-MJPR21OA).

\appendix
\begin{figure}
    \centering
    \includegraphics[width=0.75\textwidth]{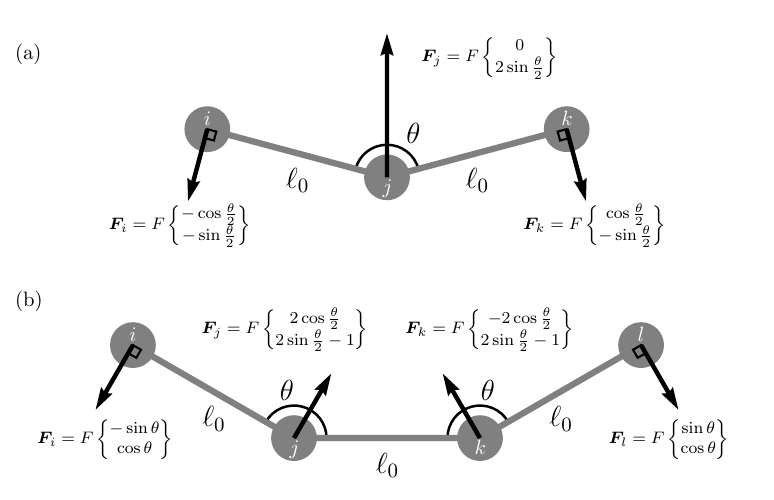}
    \caption{Schematics of structures with (a) three beads with a single joint, and (b) the four beads with two joints. \label{fig:threebeads_joint}}
\end{figure}

\section{Relaxation time for bending}
Consider a simple joint structure with three spheres (beads $i$, $j$ and $k$) or four spheres (beads $i$, $j$, $k$ and $l$) as shown in Fig.~\ref{fig:threebeads_joint}.
In this appendix, we derive an analytical solution of the time evolution of  \rev{the} angle $\theta(t)$ under \rev{the influence of a combined passive and active torque} where $0 \leq \theta(t) \leq \pi$.
Note that the hydrodynamic interactions between the beads are ignored, and the arm length $\ell_0$ is fixed during this derivation.

\subsection{Three beads structure}
Let us assume that the \rev{torque} is acting in the direction that the angle $\theta$ increases as Fig.~\ref{fig:threebeads_joint}(a).
\rev{The torque is realized by forcess $\bm{F}_{i,j,k}(\theta)=-\pdv*{(U_b+U_a)}{\mathbf{r}_{i,j,k}}$ acting perpendicular to the arm, as shown in the schematic, with the force strength $F=F(\theta)$ (as quantified below).}
The force- and torque-free conditions are satisfied.
By ignoring the hydrodynamic interactions, the velocities of three spheres are given as \rev{(in the following we simply use $F$)}
\begin{equation}
  \bm{v}_i = \frac{F}{\mu}
  \begin{Bmatrix}
      \displaystyle - \cos \frac{\theta}{2} \\
      \displaystyle - \sin \frac{\theta}{2}
  \end{Bmatrix}, \;\;\;
  \bm{v}_j = \frac{F}{\mu}
  \begin{Bmatrix}
      0 \\
      \displaystyle 2 \sin \frac{\theta}{2}
  \end{Bmatrix}, \;\;\;
  \bm{v}_k = \frac{F}{\mu}
  \begin{Bmatrix}
      \displaystyle \cos \frac{\theta}{2} \\
      \displaystyle - \sin \frac{\theta}{2}
  \end{Bmatrix}
\end{equation}
where $\mu$ is the simple drag coefficient $\mu = 6 \pi \eta a$.
The relative position vectors and the velocity differences can be given as follows: 
\begin{equation}
  \bm{r}_{ij} = \bm{r}_j - \bm{r}_i = 
  \ell_0
  \begin{Bmatrix}
      \displaystyle \sin \frac{\theta}{2} \\
      \displaystyle - \cos \frac{\theta}{2}
  \end{Bmatrix}, \;\;\;
  \bm{r}_{kj} = \bm{r}_k - \bm{r}_i = 
  \ell_0
  \begin{Bmatrix}
      \displaystyle - \sin \frac{\theta}{2} \\
      \displaystyle - \cos \frac{\theta}{2}
  \end{Bmatrix},
\end{equation}
\begin{equation}
  \bm{v}_{ij} = \bm{v}_j - \bm{v}_i
  = \frac{F}{\mu}
  \begin{Bmatrix}
      \displaystyle \cos \frac{\theta}{2} \\
      \displaystyle 3 \sin \frac{\theta}{2}
  \end{Bmatrix}, \;\;\;
  \bm{v}_{kj} = \bm{v}_j - \bm{v}_k
  = \frac{F}{\mu}
  \begin{Bmatrix}
      \displaystyle - \cos \frac{\theta}{2} \\
      \displaystyle 3 \sin \frac{\theta}{2}
  \end{Bmatrix}.
\end{equation}
Now, the time derivative of $\theta$ is given using the geometry as
\begin{equation}
  \dot{\theta} = - \frac{\dot{\bm{n}}_{ij} \cdot \bm{n}_{kj} + \bm{n}_{ij} \cdot \dot{\bm{n}}_{kj}}{\sqrt{1 - (\bm{n}_{ij} \cdot \bm{n}_{kj})^2}}
\end{equation}
where $\bm{n}_{ij} = \bm{r}_{ij}/\ell_0$ and $\bm{n}_{kj} = \bm{r}_{kj}/\ell_0$ are the normal vectors.
Using relations
\begin{align}
  \dot{\bm{n}}_{ij}
  &= \frac{\bm{v}_{ij}}{\ell_0} - \frac{(\bm{r}_{ij} \cdot \bm{v}_{ij}) \bm{r}_{ij}}{\ell_0^3}  
  = \frac{F}{\ell_0 \mu}
  \begin{Bmatrix}
    \displaystyle \cos \frac{\theta}{2} + \sin \theta \sin \frac{\theta}{2} \\
    \displaystyle 3 \sin \frac{\theta}{2} - \sin \theta \cos \frac{\theta}{2}
  \end{Bmatrix} \\
  \dot{\bm{n}}_{kj}
  &= \frac{\bm{v}_{kj}}{\ell_0} - \frac{(\bm{r}_{kj} \cdot \bm{v}_{kj}) \bm{r}_{kj}}{\ell_0^3}  
  = \frac{F}{\ell_0 \mu}
  \begin{Bmatrix}
    \displaystyle - \cos \frac{\theta}{2} + \sin \theta \sin \frac{\theta}{2} \\
    \displaystyle 3 \sin \frac{\theta}{2} + \sin \theta \cos \frac{\theta}{2}
  \end{Bmatrix}
\end{align}
and $\cos \theta = \bm{n}_{ij} \cdot \bm{n}_{kj}$, the time derivative of $\theta$ is given as
\begin{equation}
  \dot{\theta} = \frac{2F}{\ell_0 \mu} (2 - \cos \theta). 
\end{equation}
\rev{Since the force strength $F$  results  from a superposition of a passive bending spring of stiffness $k_b$ together with an active torque $L_\mathrm{max}$,
it is $\theta-$dependent and}
can be described as $F(\theta) = \{ k_b (\pi - \theta) - L_\mathrm{max} \}/\ell_0$, \rev{and} the differential equation is now given as
\begin{equation}
  \dot{\theta}^*
  = \frac{\ell_0^3 \eta}{L_\mathrm{max}} \dot{\theta}
  = \frac{1}{3\pi a^*} (2 - \cos \theta) \{ k_b^* (\pi - \theta) - 1 \}
\end{equation}
By defining a new angle $\phi = \pi - \theta$ and the angle is sufficiently small $\phi \ll 1$, the equation can be simplified as
\begin{equation}
  \dot{\phi}^* = -\frac{1}{\pi a^*} (k_b^* \phi - 1).
\end{equation}
Solving the equation, we finally have the following equation under an initial condition $\theta(t=0) = \pi$ as
\begin{equation}
  \theta(t) = \pi - \phi(t) = \pi - \frac{1}{k_b^*} + \frac{1}{k_b^*} \exp \qty(-\frac{k_b^*}{\pi a^*} t^*) \label{eq:phi}
\end{equation}
and the relaxation time of this bending process is $\tau^* = \pi a^*/k_b^*$

\subsection{Four beads structure}
Next, the time evolution of the angle $\theta$ is evaluated for the four beads structure, which is shown in Fig.~\ref{fig:threebeads_joint}(b).
\rev{Now we assume that the same torque $\lmax$ without a phase difference is applied to the two inner joints (beads $j$ and $k$).
Again, considering both active and passive bending,
the force strength acting on each bead is given as shown in the schematic, where again $F(\theta) = \{ k_b (\pi - \theta) - L_\mathrm{max} \}/\ell_0$, and} the time evolution of the angle $\theta$ from a flat state can be solved in the same manner as the three beads system.
The time evolution is
\begin{equation}
  \theta(t) = \pi - \phi(t) = \pi - \frac{1}{k_b^*} + \frac{1}{k_b^*} \exp \qty(-\frac{k_b^*}{3 \pi a^*} t^*).
\end{equation}
and the resultant relaxation time $\tau^* = 3 \pi a^3/k_b^*$ is three times longer than \rev{for} the three beads structure.

\bibliography{reference,references-RL}

\end{document}